\documentclass[twocolumns]{aa}

\usepackage{graphicx, multirow}
\usepackage{txfonts}
\usepackage{hyperref}
\begin{document}

   \title{Identifying Radio Active Galactic Nuclei with Machine Learning and Large-Area Surveys}

   \author{Xu-Liang Fan \inst{1,2}
          \and
          Jie Li\inst{1,2}
          }

    \institute{School of Mathematics, Physics and Statistics, Shanghai University of Engineering Science, Shanghai 201620, People's Republic of China\\
      \email{fanxl@sues.edu.cn}
      \and
     Center of Application and Research of Computational Physics, Shanghai University of Engineering Science, Shanghai 201620, People's Republic of China\\
             }

   \date{Received xxx; accepted xxx}


  \abstract
   {Active galactic nuclei (AGNs) and star forming galaxies (SFGs) are the primary sources of extragalactic radio sky. But it is difficult to distinguish the radio emission produced by AGNs from that by SFGs, especially when the radio sources are faint. \citet{2023MNRAS.523.1729B} classified the radio sources in LoTSS Deep Fields DR1 through multiwavelength SED fitting. With the classification results of them, we perform a supervised machine learning to distinguish radio AGNs and radio SFGs.}
   {We aim to provide a supervised classifier to identify radio AGNs, which can get both high purity and completeness simultaneously, and can easily be applied to datasets of large-area surveys.}
   {The classifications of \citet{2023MNRAS.523.1729B} are used as the true labels for supervised machine learning. With the cross-matched sample of LoTSS Deep Fields DR1, AllWISE and Gaia DR3, the features of optical and mid-infrared magnitude and colors, are applied to train the classifier. The performance of the classifier is evaluated mainly by the $precission$, $recall$ and $F_1$ score of both AGNs and non-AGNs. }
   {By comparing the performance of six learning algorithms, CatBoost is chosen to construct the best classifier. The best classifier get $precision = 0.974$, $recall = 0.865$ and $F_1 = 0.916$ for AGNs, $precision = 0.936$, $recall = 0.988$ and $F_1  = 0.961$ for non-AGNs. After applying our classifier to the cross-matched sample of LoTSS DR2, AllWISE and Gaia DR3, we obtain a sample of 49716 AGNs and 102261 non-AGNs. The reliability of these classification results is confirmed by comparing with the spectroscopic classification of SDSS. The $precission$ and $recall$ of AGN sample can be as high as 94.2\% and 92.3\%, respectively. We also train a model to identify radio excess sources. The $F_1$ scores are 0.610 and 0.965 for sources with and without radio excess, respectively.}
   {}


\titlerunning{Radio AGN Identification with Machine Learning}
\authorrunning{Fan \& Li}
   \maketitle
%

\section{Introduction}
Active galactic nuclei (AGNs) and star forming galaxies (SFGs) are the most primary populations of extragalactic radio sky~\citep{2016A&ARv..24...13P}. Both of them are important for understanding the evolution of galaxies and supermasssive black holes across cosmic time~\citep{2022A&ARv..30....6M}. Relativistic jets in AGNs are believed to be the most powerful extragalactic radio sources. Traditionally, these type of AGNs hosted powerful jets are classified as radio-loud (RL) AGNs, which are defined through the ratio of radio to optical emission~\citep{1989AJ.....98.1195K}. At faint radio sky, SFGs become numerically dominant over RL and radio-quiet (RQ) AGNs~\citep{2016A&ARv..24...13P, 2023MNRAS.523.1729B}.

Radio morphology is the direct method to identify AGN jets. Powerful radio lobes and jets can be observed in bright radio AGNs on the resolution of arcsecond~\citep{2019ARA&A..57..467B}. The detections of radio cores or low power jets, which might be ubiquitous in RQ AGNs~\citep{2008ARA&A..46..475H}, usually needs the usage of high sensitivity very long baseline interference observations at milliarcsecond resolution~\citep{2005ApJ...621..123U, 2023MNRAS.518...39W}. In addition to radio morphology, other radio properties, such as radio spectra and brightness temperature, are helpful to distinguish radio emission from AGNs and SFGs~\citep{2019NatAs...3..387P, 2022MNRAS.515.5758M}. To estimate these properties accurately, resolving radio cores and other radio structures is also preferred. Due to the limited angular resolution, majority of extragalactic radio sources are still unresolved\footnote{\citet{2023A&A...678A.151H} presented redshift estimations for more than half of radio sources in LoTSS DR2. Combining the redshift and source extension information in~\citet{2023A&A...678A.151H}, one can estimate the fraction of resolved radio sources for different redshift bins. The fraction decreases from 34\%, 19\% to 7\% for the redshift bins [$<$ 0.01], [0.01, 0.1] and [$>$ 0.1], respectively.} in current large-area radio surveys~\citep{2022A&A...659A...1S}. Thus, it is difficult to identify jetted AGNs in large dataset by the information from radio observations alone~\citep{2022A&ARv..30....6M}. Radio excess is then employed to identify jetted AGNs. In the literature, radio excess sources can be identified based on the deviation from radio/far-infrared (FIR) correlation~\citep{2015MNRAS.453.1079B}, from radio/mid-infrared (MIR) correlation~\citep{2022ApJ...939...26Y, 2024ApJ...966...53F} or from radio/star formation rate (SFR) correlation~\citep{2023MNRAS.523.1729B}.

Apart from radio excess, other multiwavelength features are available to separate RQ AGNs and SFGs. Optical spectroscopy and multiwavelength spectral energy distribution (SED) fitting are two of widely used methods. Quasars, galaxies or stars can be classified by template matching in spectroscopic surveys, such as SDSS~\citep{2012AJ....144..144B, 2020ApJS..250....8L}. AGNs in galaxies can then be classified by the so-called BPT diagram~\citep{1981PASP...93....5B}. Other useful spectroscopic features to identify radio AGNs include H$\alpha$ luminosity and 4000$\AA$ break strength, which depend on the SFR and specific SFR, respectively~\citep{2012MNRAS.421.1569B}. Galaxy SED fitting is a technique to fit the multiwavelength photometric data with physical models involved stellar population and dust radiation, which can derive the properties of the host galaxies. For several fitting codes, the radiation from AGNs are also considered. Thus the fraction of AGN contribution to observed SEDs can be determined (see~\citealt{2023ApJ...944..141P, 2023MNRAS.523.1729B} and references therein). In addition, infrared colors, optical colors and X-ray properties are also used to select AGN candidates~\citep{2016A&ARv..24...13P, 2022A&ARv..30....6M}.

As high-z and high-luminosity subsample of AGNs, quasars are expected to have zero parallaxes and proper motions. Astrometric measurement is thus useful to distinguish quasars from stars~\citep{2019MNRAS.489.4741S, 2023A&A...674A..41G}. The largest advantage of astrometry-based AGN classification is that it is unbiased for selecting obscured AGNs. Obscured AGNs, whose radiation from accretion disk is obscured by dust, would be missed in optical color selected AGN samples. To avoid the contamination from distant stars, the AGN candidates classified based on astrometric information still need confirmation by their radiation properties~\citep{2018A&A...615L...8H}.

Recent years, machine learning technique has been applied to astronomical data more frequently~\citep{2010IJMPD..19.1049B, 2014sdmm.book.....I, 2021AnRSA...8..493F, 2024A&C....4800851F}, such as the classifications of astronomical objects~\citep{2011ApJ...733...10R, 2016ApJS..225...31L, 2018arXiv180809955B, 2019ApJ...872..189K, 2023A&C....4400728L}, estimation of physical parameters~\citep{2010ApJ...712..511C, 2022A&C....4100646G, 2023MNRAS.518.4921L, 2024A&A...688A..33Z}, and analyses of astronomical images~\citep{2016A&A...591A..54K, 2019ApJS..240...34M, 2020MNRAS.491.2481B, 2023ApJS..268...34D}. Although there are many works focusing on AGN and quasar classifications, which can identify the feature of AGNs from photometric~\citep{2020A&A...639A..84C, 2024ApJS..271...54F}, spectroscopic~\citep{2018arXiv180809955B, 2023AJ....165..153T}, image~\citep{2019ApJS..240...34M, 2022arXiv221207881G}, or time series data~\citep{2019ApJ...881L...9F, 2024MNRAS.529.2877M}, few machine learning based classifications are performed for distinguishing radio AGNs from radio SFGs.

\citet{2023A&A...675A.159K} performed a supervised classification for radio AGNs and radio SFGs based on the multiwavelength data from~\citet{2021A&A...648A...3K}, including deep surveys from radio, infrared, to optical and ultraviolet. The classification labels used for their training were taken from~\citet{2023MNRAS.523.1729B}, which were obtained from SED fitting. For AGNs, their classifier got an $precision$ of 0.87, and an $recall$ of 0.78. They also attempted to train their classifier with different combinations of photometric bands. For the training with fewer bands, the performance for AGNs became worse. Especially, the $precision$ and $recall$ have not reached their highest values with the same training features. High $precision$ was derived when all the features including redshift were considered, while high $recall$ was favored when redshift was excluded (their Table 5). Another limitation of their classifier is that the training sample focused on several small sky area where the coverage of multiwavelength data are plentiful. Their training features from radio to ultraviolet bands is hard to collected for larger sky areas currently. Thus their classifier is hard to predict classifications on new datasets of large-area surveys.

In this paper, we create a supervised machine learning based classifier to identify radio AGNs, with true labels from the results of The Low Frequency Array (LOFAR) Two-metre Sky Survey (LoTSS) Deep Fields first data release (DR1)~\citep{2021A&A...648A...1T, 2021A&A...648A...2S, 2023MNRAS.523.1729B}, and training features from large-area surveys, i.e., AllWISE~\citep{2014yCat.2328....0C} and Gaia DR3~\citep{2023A&A...674A...1G}. Then we apply it to the cross-matched catalog of LoTSS DR2~\citep{2022A&A...659A...1S}, AllWISE and Gaia DR3. The reliability of the classifier is independently examined by comparing with the spectroscopic classifications of Sloan Digital Sky Survey (SDSS) DR17~\citep{2022ApJS..259...35A}. Section~\ref{sec:data} describes the datasets we use, as well the evaluation metrics for machine leaning. In section~\ref{sec:results}, we present our classifier with best performance. The classification results applied to unclassified large-area surveys are also given. In section~\ref{sec:dis}, the dependence on different groups of features, the reliability of our classifier, and the performance on identifying radio excess sources are discussed. Section~\ref{sec:summary} summarizes our main conclusions.

\section{Dataset and Method}\label{sec:data}
\citet{2023MNRAS.523.1729B} collected deep multiwavelength data and applied SED fitting for 81951 radio sources of LoTSS Deep Fields DR1. LoTSS Deep Fields are a deep sky survey with 6$\arcsec$ resolution and targeted noise level of $10-15 \mu$ Jy beam$^{-1}$ at about 150 MHz. It covers about 25 deg$^2$ of northern sky for its first Data release, which contains three fields (ELAIS-N1, Bo\"{o}tes, and Lockman Hole)~\citep{2021A&A...648A...1T, 2021A&A...648A...2S}. For these fields, deep multiwavelength data from ultraviolet to FIR are plentiful~\citep{2021A&A...648A...3K}. Using the multiwavelength photometric data, \citet{2023MNRAS.523.1729B} applied four different codes (named MAGPHYS, BAGPIPGS, CIGALE and AGNFITTER) to fit the SEDs of radio sources in LoTSS Deep Fields DR1. Two SED fitting codes, MAGPHYS and BAGPIPGS, only account for the emission from stars and dust inside the galaxies. Thus, they would show poor reliability when they are applied to fit AGNs. The other two, CIGALE and AGNFITTER consider the extra emission from AGNs, who can determine the AGN fraction based on the contribution from torus. Combining the estimations of AGN faction and fitting reliability of the four codes, they separated the sources with and without AGN contributions to their SEDs, i.e., AGNs and non-AGNs.

SFR can also be estimated by all the four SED fitting codes. In \citet{2023MNRAS.523.1729B}, they used the consensus value of SFR to maximize the advantage of different codes. For AGNs, the SFR estimations by CIGALE were favored. SFR estimations by AGNFITTER were used only when the fittings of CIGALE were considered as unreliable. For non-AGNs, the mean SFRs of MAGPHYS and BAGPIPGS were used. Based on the consensus SFRs, \citet{2023MNRAS.523.1729B} derived a relation between radio luminosity and SFRs, where the radio luminosities are expected from the SF activities. Then they identified radio excess sources through the offset to large radio luminosity relative to this radio/SFR relation. Combining the classification of AGNs and radio excess sources, \citet{2023MNRAS.523.1729B} classified radio sources into SFGs, RQ AGNs, low-excitation radio galaxies (LERGs) and high-excitation radio galaxies (HERGs) in LoTSS Deep Fields DR1. Sources without neither AGN contributions to their SEDs nor radio excess are classified as SFGs. Sources with AGN contributions but no radio excess are classified as RQ AGNs. Sources with radio excess can be divided into LERGs and HERGs based on the absence or presence of AGN SEDs, respectively. In this paper, we use the binary classifications of AGNs and non-AGNs\footnote{Actually, both SFGs and LERGs are included in non-AGN sample. A four-class classifier should be more suitable to distinguish radio sources with different properties. However, LERGs (61) and HERGs (91) are few in the main sample. More importantly, the four-class can be exactly inferred combining AGN classification and radio excess classification. Thus building two binary class classifiers is also adequate. We will discuss the classifier on radio excess sources in section~\ref{sec:radio}. } in~\citet{2023MNRAS.523.1729B} as the labels to train the data.

In order to identify new radio AGNs with large datasets, we prefer to select the multiwavelength photometric data of large-area surveys as the features for training. Although ultraviolet, X-ray and $\gamma$-ray observations are useful for AGN selection, the sensitivities and sample sizes of current surveys at these bands are still limited compared with those at low energy bands. Thus we only consider radio, infrared and optical surveys to construct our samples for machine learning.

Wide-field Infrared Survey Explorer (\textit{WISE}) surveys the MIR sky at 3.4, 4.6, 12, and 22$\mu$m ($W1, W2, W3$, and $W4$, respectively)~\citep{2010AJ....140.1868W}. AGNs are suggested to locate at an unique region on \textit{WISE} color-color diagram due to the external contributions from torus, which prompts several AGN selection criteria based on \textit{WISE} MIR colors~\citep{2011ApJ...735..112J, 2012ApJ...753...30S, 2012MNRAS.426.3271M, 2013ApJ...772...26A}. Optical colors are also widely used to selected AGN/quasar candidates, which are bluer for AGNs than those for stars and normal galaxies due to radiation from accretion disk of AGNs~\citep{1997iagn.book.....P, 2024A&A...691A...1E}. SDSS is one of most famous optical surveys with both multiband imaging and spectroscopy. Its photometry is performed with five filters, $u, g, r, i$ and $z$~\citep{2000AJ....120.1579Y}. $Gaia$ space mission is aimed to measure the three-dimensional spatial distribution of billions of astronomical objects with high accuracy astrometry and spectrophotometry~\citep{2016A&A...595A...1G}. Its third data release provides precise celestial positions, proper motions, parallaxes, as well as three broad band photometry in $G, BP$, and $RP$ for more than a billion sources across the entire sky~\citep{2023A&A...674A...1G}.

To construct the sample for machine learning and collect multiband photometric data points as training features, we cross match LoTSS Deep Fields DR1 with AllWISE and Gaia DR3 with \textit{TOPCAT}~\citep{2005ASPC..347...29T}. The radius of cross-matching are all set to 1$\arcsec$ to avoid spurious matches through this paper. After the cross-matching, we exclude the sources with no magnitude measurements of $W1, W2, W3, W4$, $G, BP$, and $RP$ bands\footnote{Some of machine learning algorithms can handle null values, such as RF, HGBT and CatBoost. We apply the default mode for null values of CatBoost with the 3034 objects in the original cross-matching sample, where the missing values are processed as the minimum value. The performance of the classifier changes slightly compared with our best classifier (section~\ref{sec:classifier}), with the $F_1 = 0.899$ and $0.957$ for AGNs and non-AGNs.}, as well as those with null values of magnitude errors in AllWISE. The later is for the 2$\sigma$ detection of AllWISE\footnote{https://wise2.ipac.caltech.edu/docs/release/allwise/expsup/sec2\_1a.html}. This results in a sample with 1698 sources, which contains 594 AGNs, 1101 non-AGNs, 160 sources with radio excess, and 1491 sources without radio excess. This sample (hereafter the main sample) is used for the supervised machine learning, which includes the classification labels from~\citet{2023MNRAS.523.1729B}, integrated radio flux density from LoTSS Deep Fields DR1, MIR photometric magnitude of $W1, W2, W3$ and $W4$ from AllWISE, magnitude of $G, BP, RP$ and two astrometric parameters (proper motion and parallax) from Gaia DR3. Two astrometric features are only used when their importance is evaluated in section~\ref{sec:features}.

The main sample with features and labels is then split into two parts. 70 percent of sources are used as training set, while remaining 30 percent are taken as test set. The classifier is trained on training set, while test set is used to evaluate the performance of the classifier. The performance of the classifier with machine learning can be evaluated by several metrics, such as $accuracy$, $precision$, $recall$ and $F_1$ score. The $accuracy$ score returns the fraction where predicted labels match the true labels. It is a simple indicator to evaluate the overall results of machine learning. Its value would be dominated by the majority class when the numbers of the both classes are imbalance. $Precision$ presents the ability of the classifier to predict true labels of each class, and $recall$ presents the ability of the classifier to find all the true labels of each class. $Precision$ and $recall$ correspond to the purity and completeness of a sample, respectively. The $F_1$ score is the harmonic mean of $precision$ and $recall$, where the relative contribution of $precision$ and $recall$ are equal. The definition of above metric is as follows,

\begin{subequations}
\begin{align}
         accuracy &= \frac{1}{n_{samples}} \sum_{i=0}^{n_{samples}-1} I(y_{true, i} = y_{prediction, i}) \\
         &= \frac{TP+TN}{TP+TN+FP+FN} \nonumber\\
         precision &= \frac{TP}{TP+FP} \\
         recall &= \frac{TP}{TP+FN} \\
         F_1 &= \frac{2 \times precision \times recall}{precision + recall}
         \label{equ:metrics}
\end{align}
\end{subequations}


where $I(x)$ is the indicator function. When $y_{true, i} = y_{prediction, i}$ it returns 1, otherwise it returns 0. The TP, TN, FP, and FN refer to the number of true positives, true negatives, false positives and false negatives, respectively.

In general, if two types of sources have overlapping parameter spaces, a classification criterion would be hard to achieve high purity and completeness simultaneously. Improving purity often means excluding more potential candidates, which would reduce the completeness. On the contrary, improving completeness would include candidates with low reliability. Machine learning technique may improve the situation with multidimensional features. In order to evaluate how the classifier addresses this issue, both $precision$ and $recall$ are presented in this paper. As a metric considering both $precision$ and $recall$, $F_1$ score is the ideal indicator to evaluate the overall performance of a classifier. It is thus used to derive the optimal algorithms and hyperparaters. In addition, the sample of AGNs and non-AGNs in the main sample are imbalance, thus the metrics of both classes are taken into account individually, especially the minority sample, AGNs. $Accuracy$ is only used to evaluate feature importance in section~\ref{sec:features}. Confusion matrix, where the numbers of true and false prediction for each class are visually presented, is also used to check the performance of our classifier. In the best-case scenario, the confusion matrix is expected to be purely diagonal, with non-zero elements on the diagonal, and zero elements otherwise~\citep{2019arXiv190407248B}.

The receiver operating characteristic (ROC) curve, as well as the Area Under the Curve (AUC) are applied to visually compare different algorithms of machine learning ~\citep{1997PatRe..30.1145B, 2019arXiv190407248B}. The Y-axis and X-axis of ROC curve are the true positive rate (TPR) and the false positive rate (FPR), which correspond to $recall$ of AGNs and $1 - recall$ of non-AGNs, respectively. A perfect machine learning model will get $recall = 1$ for both classes. That means AUC also equals to 1.

\section{Results of Machine Learning}\label{sec:results}
\subsection{The Classifier with Best Performance}\label{sec:classifier}
The performance of a machine learning model can be affected by several factors, such as the datasets, the features of training, the learning algorithms, and the hyperparameters of learning algorithms. We firstly compare the performance for six widely applied supervised machine learning algorithms with their default parameters, named Support Vector Machines (SVC), Multi-layer Perceptron (MLP), random forests (RF), AdaBoost, Histogram-Based Gradient Boosting (HGBT), and CatBoost. Then the hyperparameters of the optimal algorithm are tuned to determine the best classifier. The effects from datasets and training features are discussed in section~\ref{sec:features} and section~\ref{sec:reliability}.

Among these six algorithms, SVC finds a hyperplane that best separates the given classes. Sources are classified according to their location with respect to the hyperplane~\citep{2019arXiv190407248B}. MLP is a neural network, which can consist multiple non-linear hidden layers between the input and the output layer. The other four algorithms are ensemble models, which combine the predictions of the base estimators in order to improve the performance over a single estimator. RF is a bagging ensemble method, where the ensemble consists of a diverse collection of base estimators that are independently trained on different subsets of the data. The other three, AdaBoost, HGBT, and CatBoost are boosting ensembles. In which the base estimators are built sequentially. AdaBoost improves the model performance through increasing the weights of the bad learned data, while HGBT and CatBoost make it by optimizing the gradient of the loss function. Different approaches are employed by HGBT and Catboost to overcome the disadvantage of gradient boosting algorithm. HGBT use a histogram-based approach to speed up the training process on large datasets, while CatBoost can reduce overfitting through an efficient strategy to build tree structure~\citep{2018arXiv181011363V}.

The six algorithms are running with  \textit{scikit-learn}\footnote{https://scikit-learn.org/stable/index.html}~\citep{2011JMLR...12.2825P} and \textit{CatBoost}\footnote{https://catboost.ai/}~\citep{2017arXiv170609516P, 2018arXiv181011363V} in \textit{python}. The default hyperparameters of each algorithm are listed in Table~\ref{tab:algorithm}. The training features include all the optical and MIR magnitudes and colors ($W1mag$, $W2mag$, $W3mag$, $W4mag$, $W1-W2$, $W2-W3$, $W3-W4$, $Gmag, BPmag, RPmag, BP-G, G-RP$).

Table~\ref{tab:algorithm} lists the metrics evaluated from the test set of our main sample for each algorithm, where $precission$, $recall$, and $F_1$ score of AGNs and non-AGNs are presented separately. Generally, four ensemble algorithms show better performance than SVM and MLP. CatBoost has highest metrics among three boosting algorithms. RF shows the best $precision$ of AGNs and $recall$ of non-AGNs. The biggest difference of the metrics comes from the $recall$ and $F_1$ score of AGNs. For both metrics, Catboost has the best performance. In addition, Catboost shows both the highest $F_1$ score for AGNs and non-AGNs among six algorithms. Thus, CatBoost is favored to construct the best classifier.

\begin{table*}
\centering
\small
\caption{Hyperparameters and model performance for different machine learning algorithms\label{tab:algorithm}}
\begin{tabular}{cccccccc}
\hline\hline
\multirow{2}{*}{Algorithm} & \multirow{2}{*}{Hyperparameters} & \multicolumn{3}{c}{AGN} & \multicolumn{3}{c}{non-AGN} \\
& & $precision$ & $recall$ & $F_1$ score & $precision$ & $recall$ & $F_1$ score\\
\hline
 \multirow{2}{*}{SVC} & $C = 1.0,~kernel = 'rbf',$ & \multirow{2}{*}{0.976} & \multirow{2}{*}{0.718} & \multirow{2}{*}{0.827} & \multirow{2}{*}{0.875} & \multirow{2}{*}{0.991} & \multirow{2}{*}{0.929} \\
 & $gamma = 'scale'$ &&&&&&\\
 &  &&&&&&\\
 \multirow{2}{*}{MLP} & $solver = 'adam',~alpha = 0.0001,$ & \multirow{2}{*}{0.959} & \multirow{2}{*}{0.829} & \multirow{2}{*}{0.890} & \multirow{2}{*}{0.920} & \multirow{2}{*}{0.982} & \multirow{2}{*}{0.950} \\
 & $hidden\_layer\_sizes = (100,)$ &&&&&&\\
 &  &&&&&&\\
 \multirow{4}{*}{RF} & $n\_estimators = 100,~max\_features = 'sqrt',$ & \multirow{4}{*}{0.986} & \multirow{4}{*}{0.847} & \multirow{4}{*}{0.911} & \multirow{4}{*}{0.928} & \multirow{4}{*}{0.994} & \multirow{4}{*}{0.960} \\
 & $max\_samples = None,~min\_samples\_split = 2, $ &&&&&&\\
 & $bootstrap = True,~max\_leaf\_nodes = None, $ &&&&&&\\
 & $max\_depth = None $ &&&&&&\\
 &  &&&&&&\\
 \multirow{2}{*}{AdaBoost} & $n\_estimators = 50,~learning\_rate = 1.0, $ & \multirow{2}{*}{0.947} & \multirow{2}{*}{0.847} & \multirow{2}{*}{0.894} & \multirow{2}{*}{0.927} & \multirow{2}{*}{0.976} & \multirow{2}{*}{0.951} \\
 & $estimator = None $ &&&&&&\\
 &  &&&&&&\\
 \multirow{4}{*}{HGBT}  & $max\_iter = 100, max\_leaf\_nodes = 31,$ & \multirow{4}{*}{0.973} & \multirow{4}{*}{0.853} & \multirow{4}{*}{0.909} & \multirow{4}{*}{0.931} & \multirow{4}{*}{0.988} & \multirow{4}{*}{0.959} \\
 & $max\_depth = None, min\_samples\_leaf = 20,$ &&&&&&\\
 & $max\_bins = 255, l2\_refularization = 0.0,$ &&&&&&\\
 & $early\_stopping = 'auto'$ &&&&&&\\
 &  &&&&&&\\
 \multirow{2}{*}{CatBoost (default)} & $iterations = 1000$  & \multirow{2}{*}{0.974} & \multirow{2}{*}{0.865} & \multirow{2}{*}{0.916} & \multirow{2}{*}{0.936} & \multirow{2}{*}{0.988} & \multirow{2}{*}{0.961} \\
 & $learning\_rate = 0.01108,~depth = 6$ &&&&&&\\
 &  &&&&&&\\
 \multirow{2}{*}{CatBoost (optimal)} & $iterations = 184~(10, 2000)\tablefootmark{*}$ & \multirow{2}{*}{0.981} & \multirow{2}{*}{0.888} & \multirow{2}{*}{0.932} & \multirow{2}{*}{0.946} & \multirow{2}{*}{0.991} & \multirow{2}{*}{0.968} \\
 & $learning\_rate = 0.317~(0.001, 1)\tablefootmark{*}, $ &&&&&&\\
 & $depth = 6~(1, 16)\tablefootmark{*}$ &&&&&&\\
 \hline
\end{tabular}
\tablefoot{The lower and upper limits for hyperparameter tuning are presented in brackets.}
\end{table*}

\begin{figure}
\centering
\includegraphics[angle=0,scale=0.6]{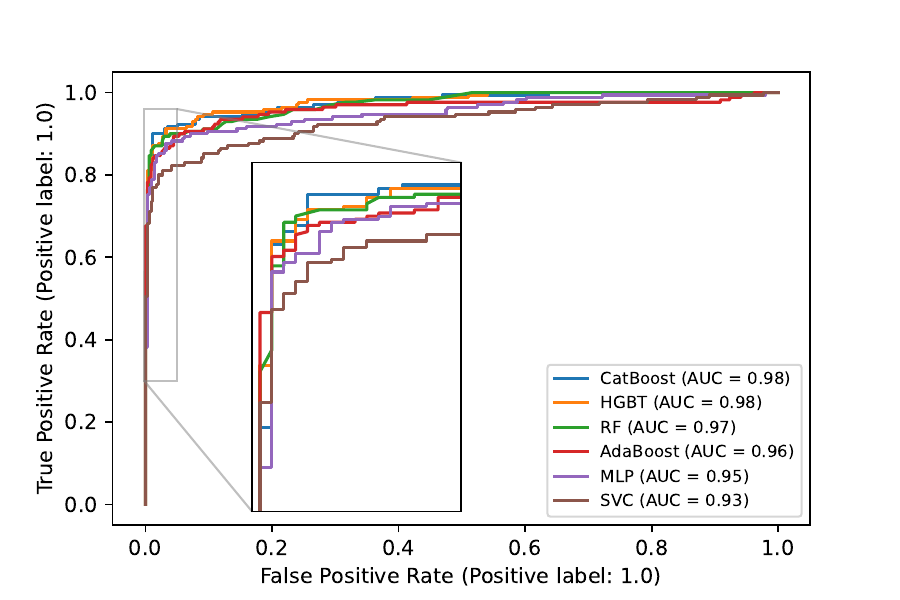}
  \caption{ROC curves of six different machine learning algorithms. The inset zooms in the region from 0 to 0.05 of FPR. The AUC of each algorithm is labeled in the legend.}
     \label{roc}
\end{figure}

Figure~\ref{roc} shows ROC curves of six algorithms, where the legend shows AUC of them~\citep{1997PatRe..30.1145B}. All the six algorithms get high TPRs when FPRs are low, which indicates they all work well for our main sample. Among them, CatBoost and HGBT show highest AUC than other algorithms. The $accuracy$ of HGBT for the training set is 1, which may indicate an overfitting problem of it\footnote{Overfitting problem of HGBT can be reduced by enabling early stopping or by reducing the maximum number of iterations. With $max\_iter = 30$, the $accuracy$ for the training set decreases to 0.981. Meanwhile, a higher $F_1$ score of 0.916 is derived, which is close to the results of CatBoost. As we just select the optimal algorithm with its default parameters, CatBoost is chosen to construct the best classifier in the following section.}. CatBoost is again favored by the ROC curve as the best algorithm to construct the best classifier.

In order to derive the best classifier, we employ the Bayesian optimization\footnote{https://github.com/bayesian-optimization/BayesianOptimization} method to tune the hyperparameters of CatBoost. The hyperparameters and the parameter spaces used for the optimizer are listed in Table~\ref{tab:algorithm}. $F_1$ score for AGNs is taken as the indicator to determine the optimal hyperparameter. The optimizer runs 100 iterations. The optimal $F_1$ score is as high as 0.932. Other metrics and the optimal values of tuned hyperparameters are listed in Table~\ref{tab:algorithm}.

\begin{figure}
\centering
\includegraphics[angle=0,scale=0.4]{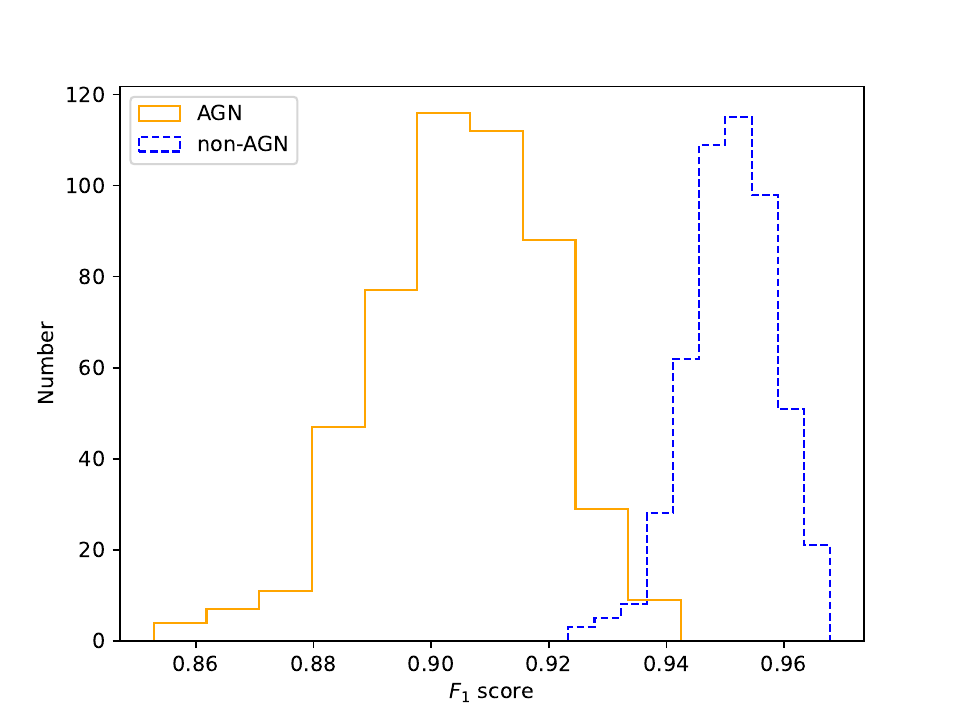}
\includegraphics[angle=0,scale=0.4]{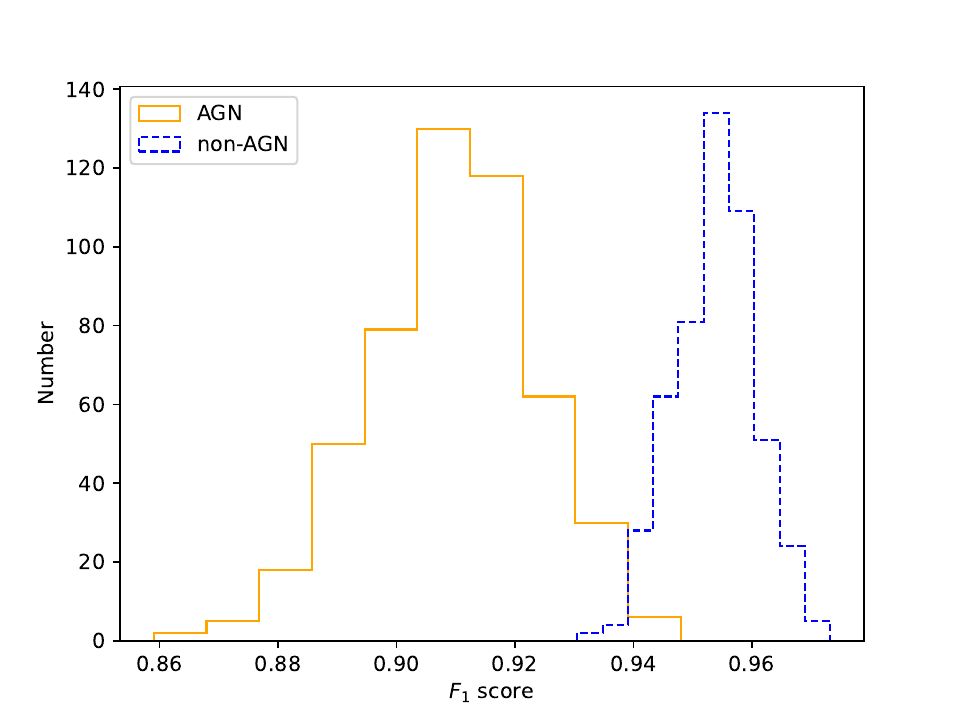}
  \caption{Distribution of $F_1$ scores for AGNs and non-AGNs with randomly splits of training set and test set. Top panel is derived based on the training of CatBoost with tuned hyperparameters, while bottom panel is derived based on the training of CatBoost with default hyperparameters.}
     \label{split}
\end{figure}

After the hyperparameter tuning, we examine the stability of model performance due to the randomly split of training and test sets. The splits of training/test set are randomly performed for 500 times. For each time, the training set is trained by CatBoost with the optimal hyperparameters. $F_1$ scores of both classes are then evaluated from the test set. The distributions of $F_1$ scores for AGNs and non-AGNs are plotted on the top panel of Figure~\ref{split}. The average $F_1$ scores for AGNs and non-AGNs are $0.905\pm0.015$ and $0.951\pm0.008$, respectively. The tuned optimal $F_1$ score 0.932 lies at the high end of the distribution for AGNs. As a comparison, the distributions of $F_1$ scores for CatBoost with default parameters are plotted on the bottom panel of Figure~\ref{split}. The average $F_1$ scores are $0.910\pm0.014$ and $0.954\pm0.007$ for AGNs and non-AGNs, respectively. Small standard deviations of the $F_1$ score distributions indicate that randomly splits of training/test set have little influence on the model training. However, the hyperparameter tuning might just select hyperparameters with high $F_1$ score by chance. The average $F_1$ score derived with default hyperparameters is even better than that derived with tuned hyperparameters. Thus in following sections, we choose CatBoost with default hyperparameters as the best model.

\begin{figure}
\centering
\includegraphics[angle=0,scale=0.55]{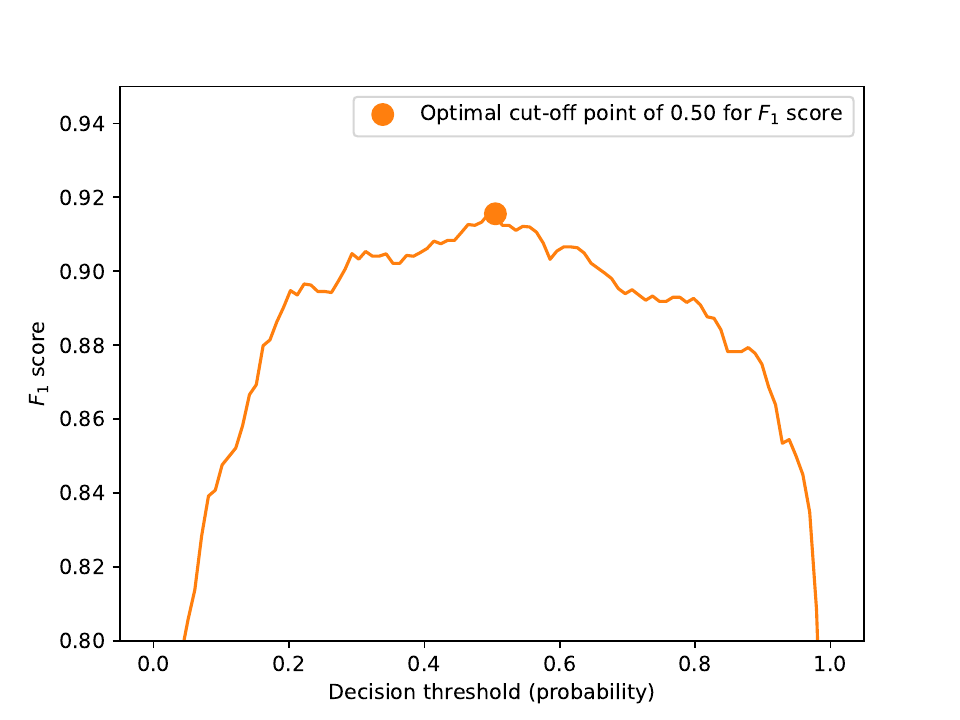}
  \caption{Variation of $F_1$ scores for AGNs along with classification thresholds. The model reaches the best $F_1$ score 0.916 when the threshold is 0.5.
          }
     \label{threshold}
\end{figure}

The threshold of the classification also tuned with \textit{TunedThresholdClassifierCV} in \textit{scikit-learn}, which is set as $predict\_proba$ for the binary classification in our case. \textit{TunedThresholdClassifierCV} is done by maximize a metric of the classifier. $F_1$ score for AGNs is again taken as the scorer to determine the optimal threshold. Figure~\ref{threshold} shows the dependence of $F_1$ score on the classification threshold. When the threshold is 0.5, $F_1$ score reaches the highest value 0.916. Thus the default $predict\_proba = 0.5$ is used through this paper. Table~\ref{tab:matrix} shows the confusion matrix of the test set for the best classifier.

\begin{table*}
\centering
\caption{Confusion matrix for test set of best classifier \label{tab:matrix}}
\begin{tabular}{cc|ccc}
\hline\hline
& & \multicolumn{3}{c}{Predicted Labels}\\
 \cline{3-5}
& & AGN & non-AGN & $recall$\\

\hline
\multirow{3}{*}{True Labels} \vline &
AGN & 147 & 23 & 0.865\\
\hspace{47.7pt} \vline & non-AGN & 4 & 335 & 0.988\\
\hspace{47.7pt} \vline & $precision$ & 0.974 & 0.936 & \\
 \hline
\end{tabular}
\end{table*}

\subsection{Application to Unclassified Large-Area Surveys}\label{sec:application}
The features of our best classifier are all from AllWISE and Gaia DR3, thus it can be easily applied to predict large samples of AGNs and non-AGNs. As we aim to identify radio AGNs, we need an large-area survey at radio band. LoTSS DR2 uses the same High Band Antenna observations of LOFAR with LoTSS Deep field. It has similar angular resolution but a wide sky area compared with LoTSS Deep field, and will survey the entire northern sky in the future~\citep{2017A&A...598A.104S, 2021A&A...648A...1T, 2022A&A...659A...1S}. We cross match LoTSS DR2 with AllWISE and Gaia DR3 with 1$\arcsec$ separation. The obtained 151977 sources with all the training features in the cross-matched sample are classified by our best classifier (hereafter prediction set). 49716 sources are classified as AGNs, and remaining 102261 sources are classified as non-AGNs.

\begin{figure}
\centering
\includegraphics[angle=0,scale=0.4]{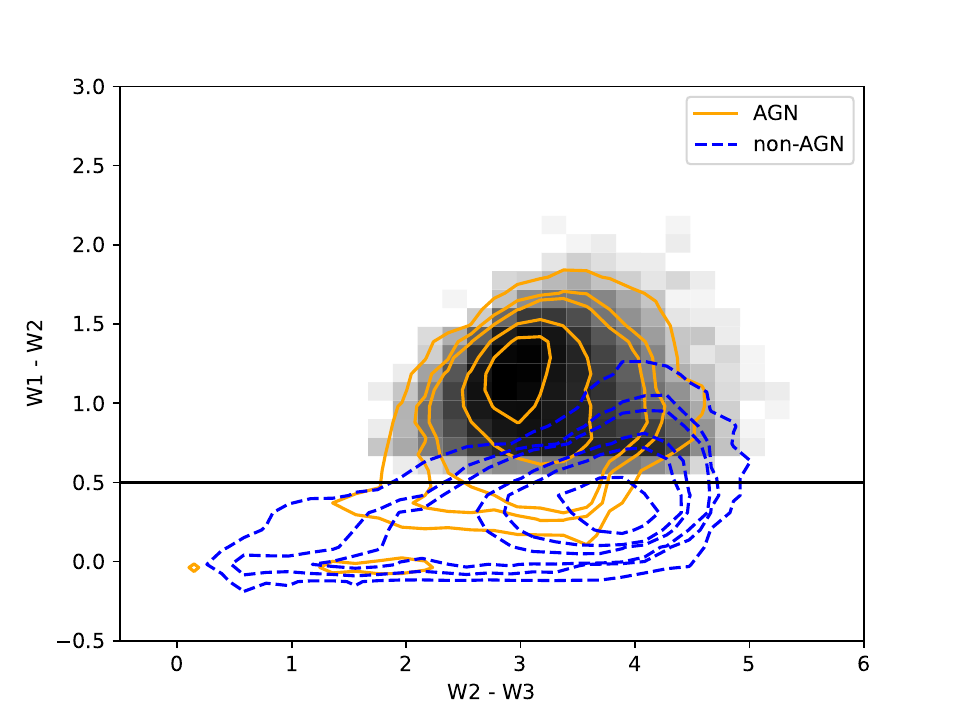}
\includegraphics[angle=0,scale=0.4]{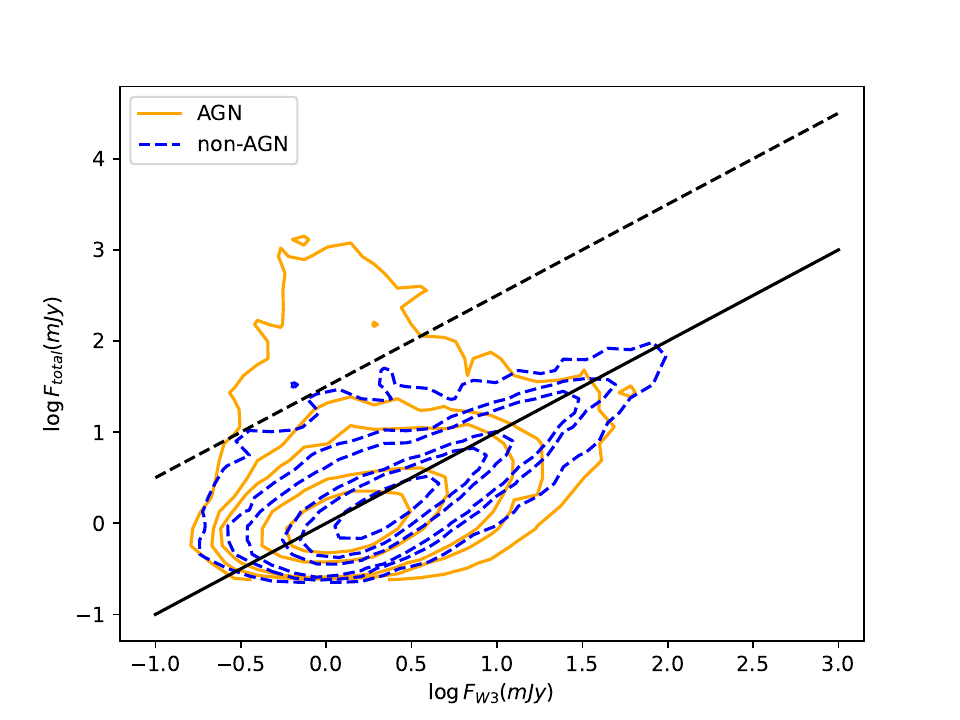}
  \caption{Top panel: \textit{WISE} color-color diagram for prediction set. The horizontal line presents the simple $W1-W2$ criterion to select AGNs, while the grey density plots show the sources satisfying the criterion combining $W1-W2$ color and $W2mag$ (see the text for details). Bottom panel: radio/MIR flux diagram for prediction set. The solid line shows the one-to-one correlation. The dashed line presents a division line to identify radio/MIR excess. For both panels, the orange contours represent AGNs. The levels of the contours are [10, 50, 100, 500, 1000]. The blue contours represent non-AGNs. The levels of the contours are [10, 50, 100, 500, 1000, 3000].
          }
     \label{prediction}
\end{figure}

The \textit{WISE} color-color diagram and radio/MIR excess are two of the most widely used methods to select radio AGNs~\citep{2025MNRAS.539.1856H}. In Figure~\ref{prediction}, the \textit{WISE} color-color diagram and radio/MIR flux diagram\footnote{The behaviors of the radio/MIR flux diagram for all four \textit{WISE} bands are similar, except that the overall MIR flux density increases from $W1$ to $W4$ band. As MIR emission of $W3$ band is usually suggested to distinguish radio AGNs from radio SFGs~\citep{2024ApJ...966...53F, 2025MNRAS.539.1856H}, or as an indicator of SFRs~\citep{2017ApJ...850...68C} in the literature, only the plot between radio and $W3$ flux density is shown in this paper.} of the prediction set are shown. The $W3$ flux density is converted from magnitude with the zero-magnitude flux density of 171.787Jy~\citep{2011ApJ...735..112J}. The predicted AGNs and non-AGN galaxies generally locate distinct regions on the \textit{WISE} color-color diagram. AGNs show redder $W1 - W2$ than non-AGNs, which is followed the expectation of the emission from AGN torus. Non-AGNs occupy the regions from elliptical galaxies to SFGs~\citep{2011ApJ...735..112J}.

There are also fractional sources overlapped between AGNs and non-AGNs. Although a high completeness can be easy to achieve with a bluer $W1 - W2$ selection criterion. Extra features besides \textit{WISE} colors are needed to improve the reliability of AGN section. \citet{2018ApJS..234...23A} proposed two criteria to identify AGNs, which can reach high reliability and high completeness, respectively. With simple criterion of $W1-W2 > 0.5$, they get 90\% completeness of AGN selection (C90 catalog). By combining $W1-W2$ color and $W2$ brightness, the reliability can reach 90\% (R90 catalog). For fainter source with $W2mag > 13.86$, the criterion is $W1-W2 > 0.650*\exp(0.153*(W2 - 13.86)^2)$. For brighter sources with $W2mag < 13.86$, the criterion is $W1-W2 > 0.650$. The C90 criterion is plotted with a black horizontal line in the top panel of Figure~\ref{prediction}. AGNs classified with R90 criterion in the prediction set are shown as the grey density plot. If the classification of our classifier is taken as true label, the $precision$ and $recall$ of these two criteria in \citet{2018ApJS..234...23A} can be estimated. The results are shown in Table~\ref{tab:mir_class}. As a comparison, the results for that the main sample is treated as the true label are also shown. The good completeness of C90 criterion is confirmed by the SED classification (0.891) and our classifier (0.949). The high $precision$ of R90 criterion for the main sample (0.936) and prediction set (0.908) indicates our classifier can achieve similar reliability on AGN selection with SED and MIR based classifications. Compared with the MIR based criteria, our classifier can reduce the dilution of AGNs from ultraluminous infrared galaxies, and improve the completeness due to the AGNs with bluer $W1 - W2$ colors.

\begin{table*}
\centering
\caption{Classification performance of two MIR color based criteria. \label{tab:mir_class}}
\begin{tabular}{ccccc}
\hline\hline
 \multirow{2}{*}{Metric} & \multicolumn{2}{c}{Main Sample} & \multicolumn{2}{c}{Prediction Set}\\
 & C90 & R90 & C90 & R90 \\
 \hline
 $precision$ & 0.714 & 0.936 & 0.675 & 0.908 \\
 $recall$ & 0.891 & 0.759 & 0.949 & 0.855 \\
 \hline
\end{tabular}
\end{table*}

Non-AGNs in the bottom panel of Figure~\ref{prediction} locate around the line that radio flux density equals to $W3$ flux density, which are similar with the known SFGs of LoTSS DR2~\citep{2024ApJ...966...53F}. Most of AGNs also show nearly equal radio flux density with $W3$ flux density, while a small fraction of AGNs exhibit obvious radio excess. For radio sources with $F_{total} < F_{W3} + 1.5$ (dashed line in the bottom panel of Figure~\ref{prediction}), AGNs and non-AGNs are generally undistinguishable on this plot. This indicates that the AGN selection method through radio/MIR excess would only select strong jetted AGNs but miss most of potential radio AGNs.

\section{Discussions}\label{sec:dis}
\subsection{Dependence on Multiwavelength Features} \label{sec:features}
\begin{figure}
\centering
\includegraphics[angle=0,scale=0.55]{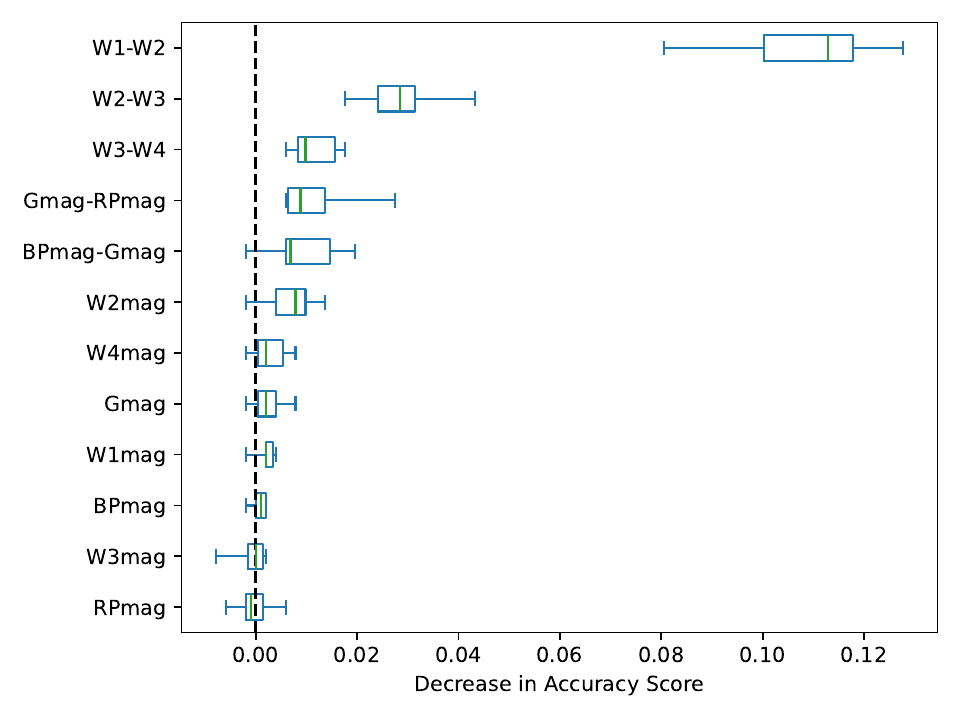}
  \caption{Importance rank with permutation feature importance method.
          }
     \label{importance}
\end{figure}

Figure~\ref{importance} shows the importance of training features with permutation feature importance method~\citep{2001MachL..45....5B}, which is defined through comparing the $accuracy$ of the test sample by permutating the features. $W1-W2$ is the most important feature to identify radio AGNs, which means that permuting $W1-W2$ will lead to most decrease in $accuracy$ of the classifier. It is expected as $W1-W2$ is widely used solely to identify AGNs. Besides $W1-W2$, other MIR colors ($W2-W3$ and $W3-W4$) are also more important than optical colors to the classifier. Optical and MIR magnitudes seems unimportant to identify radio AGNs.

The performance with different combinations of optical/MIR magnitude and colors are also compared through the metrics ($precission$, $recall$ and $F_1$ score) of each class (Table~\ref{tab:feature}). Again the differences of different combinations mainly focus on $recall$ and $F_1$ score of AGNs. The performance if only MIR colors are considered is worse than that performed with MIR magnitudes. The performance improves when both MIR colors and MIR magnitudes are taken into account. Optical colors and magnitudes show worse performance than MIR, even when MIR colors are combined with optical colors. Combining optical and MIR magnitudes achieves the second-best performance, which is next only to our best classifier utilizes all features.

The SED fitting methods classify AGNs mainly through identifying the features of torus emission at IR~\citep{2023MNRAS.523.1729B}. The unique MIR colors of AGNs are also suggested to be caused by the torus emission at different redshift~\citep{2010ApJ...713..970A}. Thus MIR features are no doubt to be important to identify AGNs in machine learning. On the other side, optical colors tend to identify unobscured AGNs. Obscured AGNs, who would be missed by optical features, can still be selected by MIR features. These two reasons result in the worse performances with optical features than those with MIR features.

In addition, we examine the performance of the classifier for several extra features, including the radio properties and astrometric parameters (Table~\ref{tab:feature}). For radio properties, we consider both the integrated radio flux density and the ratio between radio and MIR flux density. The performance has no improvement for both features. The $recall$ of AGNs decreases slightly when radio features are added. Some objects have no measurement of parallax and proper motion in the main sample. After removing the sources without these two astrometric parameters, there are 574 sources left, including 455 AGNs and 116 non-AGNs. AGNs dominates the counts of this subsample, and show better performance than non-AGNs unsurprisingly.

\begin{table*}
\small
\centering
\caption{Performance for different combinations of training features\label{tab:feature}}
\begin{tabular}{ccccccc}
\hline\hline
\multirow{2}{*}{Features} & \multicolumn{3}{c}{AGN} & \multicolumn{3}{c}{non-AGN} \\
& $precision$ & $recall$ & $F_1$ score & $precision$ & $recall$ & $F_1$ score \\
\hline
 $W1, W2, W3, W4$ & 0.978 & 0.794 & 0.877 & 0.906 & 0.991 & 0.946 \\
 $W1-W2, W2-W3, W3-W4$ & 0.943 & 0.776 & 0.852 & 0.897 & 0.976 & 0.935 \\
 $W1, W2, W3, W4, W1-W2, W2-W3, W3-W4$ & 0.965 & 0.812 & 0.882 & 0.913 & 0.985 & 0.948 \\
 $G, BP, RP, BP-G, G-RP$ & 0.901 & 0.747 & 0.817 & 0.883 & 0.959 & 0.919 \\
 $W1-W2, W2-W3, W3-W4, BP-G, G-RP$ & 0.958 & 0.812 & 0.879 & 0.912 & 0.982 & 0.946 \\
 $W1, W2, W3, W4, G, BP, RP$ & 0.966 & 0.824 & 0.889 & 0.918 & 0.985 & 0.950 \\
 $W1, W2, W3, W4, W1-W2, W2-W3, W3-W4,$ & \multirow{2}{*}{0.973} & \multirow{2}{*}{0.853} & \multirow{2}{*}{0.910} & \multirow{2}{*}{0.931} & \multirow{2}{*}{0.988} & \multirow{2}{*}{0.959} \\
 $G, BP, RP, BP-G, G-RP, F_{total}$ &&&&&&\\
 $W1, W2, W3, W4, W1-W2, W2-W3, W3-W4,$  & \multirow{2}{*}{0.973} & \multirow{2}{*}{0.841} & \multirow{2}{*}{0.902} & \multirow{2}{*}{0.925} & \multirow{2}{*}{0.988} & \multirow{2}{*}{0.956} \\
 $G, BP, RP, BP-G, G-RP, F_{total}/F_{W3}$ &&&&&&\\
 $W1, W2, W3, W4, W1-W2, W2-W3, W3-W4,$  & \multirow{2}{*}{0.963} & \multirow{2}{*}{0.949} & \multirow{2}{*}{0.956} & \multirow{2}{*}{0.816} & \multirow{2}{*}{0.861} & \multirow{2}{*}{0.838} \\
 $G, BP, RP, BP-G, G-RP, parallax, proper~motion$\tablefootmark{*} &&&&&&\\
 \hline
\end{tabular}
\tablefoot{The number of AGNs and non-AGNs containing measurements of parallax and proper motion are 455 and 116, respectively.}
\end{table*}

\subsection{Reliability of the Classifier}\label{sec:reliability}
As mentioned in section~\ref{sec:features}, \textit{WISE} colors show the largest importance on our best classifier. Thus we perform a similar learning on the cross-matched sample between LoTSS Deep Fields DR1 and AllWISE with only MIR features ($W1mag$, $W2mag$, $W3mag$, $W4mag$, $W1 -W2$, $W2 -W3$ and $W3 -W4$). The cross-matched sample contains 1429 AGNs and 5142 non-AGNs. The $precision$, $recall$ and $F_1$ score for AGNs are 0.794, 0.776 and 0.785, respectively. The performance is much worse than that for the main sample with same features (Table~\ref{tab:feature}). Although colors of Gaia and the astrometric parameters seem less important to the best classifier, the Gaia detection might change the distributions of features severely after the cross-match with Gaia DR3. In order to clarify whether this effect would affect our classification results, we perform two independent tests.

Firstly, we examine the reliability of the classification for the prediction set in section~\ref{sec:application}. Apart from SED fitting, the optical spectroscopy can also be used to classify quasars, AGNs, SFGs, quiescent galaxies, and stars~\citep{2012AJ....144..144B, 2020ApJS..250....8L}. Therefore, we use the classifications in SDSS DR17 to help us examining our classification results. Except AGNs, SDSS classified sources as galaxies and stars based on their spectra\footnote{https://www.sdss4.org/dr17/spectro/catalogs/}. When the spectroscopic objects in SDSS have \textit{class} of "QSO", or \textit{class} of "GALAXY" and subclass containing "AGN", they are treated as AGNs. Other sources with \textit{class} of "GALAXY" are treated as galaxies. Stars are objects with "STAR" classifications. In SDSS DR17, there are 34050 AGNs, 38017 galaxies and 99 stars in our prediction set.

\begin{figure}
\centering
\includegraphics[angle=0,scale=0.55]{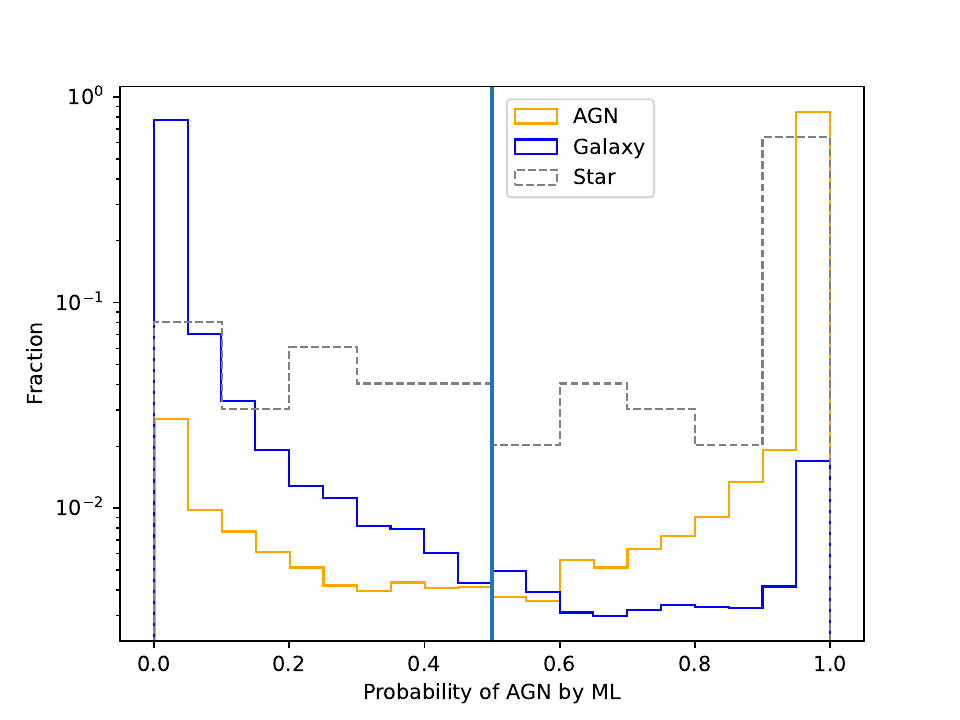}
  \caption{Histogram of predicted probability of AGN for common sources of prediction set and SDSS DR17. The orange, blue, and grey lines represent AGNs, galaxies and stars, respectively, which are classified by SDSS. The vertical line labels the critical probability (0.5) to distinguish AGNs and non-AGNs.}
     \label{com_sdss}
\end{figure}

Then we compare the classification of machine learning with that of SDSS spectroscopy for the common sources. Figure~\ref{com_sdss} shows histogram of the prediction probability of our best classifier. The orange, blue, and grey lines represent AGNs, galaxies and stars of SDSS classification, respectively. Most SDSS AGNs have AGN probabilities in the range 0.9 to 1.0, while galaxies have low AGN probabilities (0.0 to 0.1). In table~\ref{tab:sdss}, we list the confusion matrix of prediction set where the SDSS classifications are treated as true labels. The purity of AGN sample can be up to 94.2 percent, while the completeness is 92.3 percent.

\begin{table*}
\centering
\caption{Confusion matrix when compared with SDSS classifications \label{tab:sdss}}
\begin{tabular}{cc|ccc}
\hline\hline
& & \multicolumn{3}{c}{Predicted Labels}\\
 \cline{3-5}
& & AGN & non-AGN & $recall$\\

\hline
\multirow{4}{*}{True Labels} \vline &
AGN & 31442 & 2608 & 0.923\\
\hspace{47.7pt} \vline & Galaxy & 1870 & 36147 & 0.951\\
\hspace{47.7pt} \vline & Star & 74 & 25 & ...\\
\hspace{47.7pt} \vline & $precision$ & 0.942 & 0.932 & \\
 \hline

\end{tabular}
 \tablefoot{We ignore the number of stars when calculating the $precision$ and $recall$ of non-AGNs.}
\end{table*}

In addition, we attempt to build similar distributions (e.g., average magnitudes) of the training features between the main sample and the cross-matched sample of LoTSS Deep Fields DR1 and AllWISE. The average \textit{W2mag} of the main sample is 14.20, which is brighter than 15.09 of the cross-matched sample between LoTSS Deep Fields DR1 and AllWISE. We apply a simple brightness cut with $W2mag < 14.5$\footnote{This criterion is firstly determined in order to match the average $W2mag$ of the main sample. When the criterion is taken as 15.0, the average \textit{W2mag} is 14.23, which is closest to 14.20 of the main sample. The classifier gets a $F_1$ score 0.849 for AGNs. Then we attempt to maximize the $F_1$ score. The criterion changes from 14.4 to 15.1 with a step of 0.1 mag. $F_1$ score for AGNs shows the lowest value of 0.847 at 15.1 mag, and shows the highest value of 0.900 at 14.5 mag.} to the cross-matched sample of LoTSS Deep Fields DR1 and AllWISE, which results in a sample of 505 AGNs and 1164 non-AGNs with average \textit{W2mag} = 13.85. The proportion between AGNs and non-AGNs is similar to that of the main sample (594:1101). CatBoost is applied to model the MIR features of this new sample. The performance of the classifier becomes much better on this sample, with the $precision = 0.959$, $recall = 0.848$ and $F_1 =  0.900$ for AGNs.

We also apply similar magnitude cut onto the cross-matched sample between LoTSS Deep Fields DR1, AllWISE and SDSS, which contains 1106 AGNs and 4413 non-AGNs. The limiting magnitude of Gaia $G$ band is about 21~\citep{2016A&A...595A...1G, 2021A&A...649A...5F}, which is about 2 mag brighter than that of SDSS \textit{g} band~\citep{2000AJ....120.1579Y}. The average \textit{g} and \textit{W2} band magnitudes of the cross-matched sample between LoTSS Deep Fields DR1, AllWISE and SDSS are 20.54 and 14.99, respectively. We apply a magnitude cut with \textit{gmag} less than 19\footnote{This criterion is determined to match the difference on the limiting magnitude between Gaia and SDSS surveys, which is about 2 magnitudes on the mean magnitude. For a criterion of 19.5, the average \textit{g} band magnitude is 18.42, which is closest to target value of 18.54. The $F_1$ score for AGNs is also maximized by changing the criterion from 18.5 to 20.0 with a step of 0.5 mag. The best $F_1$ score is 0.886 for the criterion of 19.0, while the worst one is 0.756 for the criterion of 18.5.}, which results in the average \textit{g} and \textit{W2} band magnitudes brightening to 18.04 and 13.94. With the training features of MIR and optical magnitudes and colors ($W1mag$, $W2mag$, $W3mag$, $W4mag$, $W1 - W2$, $W2 - W3$, $W3 - W4$, $umag$, $gmag$, $rmag$, $imag$, $zmag$, $u - g$, $g - r$, $r - i$, $i - z$), CatBoost is again applied to the samples before and after magnitude cut. The $precision$, $recall$ and $F_1$ score increase to 0.969, 0.816 and 0.886 for AGNs, compared with 0.922, 0.742 and 0.822 for the sample without magnitude cut, respectively. As the sample is severely imbalanced after magnitude cut (157 AGNs versus 980 non-AGNs), we apply the Synthetic Minority Oversampling Technique (\textit{SMOTE}) to optimize the above model with a relatively balanced training sample. For imbalanced dataset, the metrics would be dominated by the majority class. \textit{SMOTE} handles this though over-sampling the minority class\citep{2011arXiv1106.1813C}. The number of AGNs and non-AGNs are set to 1200 and 2000, respectively. The $F_1$ score for AGNs increases from 0.886 to 0.909 with $precision = 0.897$ and $recall = 0.921$.

The simple magnitude cut on the above two samples shows that the distributions of training features are important to construct a well-performed classifier. Cross-matching with Gaia DR3 achieves a similar effect by selecting bright sources. Although this may change the distributions of training features, our classifier is still proven to be robust through the comparison with the spectroscopic classification of SDSS.

\subsection{Classification on Radio Excess Sources}\label{sec:radio}
Radio excess sources, whose radio emission is believed to be dominated by jets of AGNs, are usually identified through the excess of the radio emission excepted by SF activities~\citep{2015MNRAS.453.1079B, 2023MNRAS.523.1729B, 2024ApJ...966...53F}. There are few attempts to identify radio excess sources with machine learning in the literature~\citep{2023A&A...679A.101C}. For the main sample, we perform a supervised machine learning to identify radio excess sources. The true labels are also taken from~\citet{2023MNRAS.523.1729B}, which are classified based on the SFR estimated from SED fitting. For training features, two extra radio features $F_{total}$ and $F_{total}/F_{W3}$ are added together with optical and MIR magnitudes and colors. CatBoost with default parameter is applied for training. Compared with the AGN classifier, classifying radio excess sources is more difficult. The $precision$, $recall$ and $F_1$ score for radio excess sources are 0.714, 0.532 and 0.610, while they are 0.952, 0.978 and 0.965 for sources without radio excess, respectively.

\citet{2023A&A...679A.101C} performed a machine learning based classification for radio AGNs with features of optical and MIR photometries. Their strategy was different from that in our work. Their model was trained to clarify whether an AGN can be radio detected, while the AGN sample was also derived from a machine learned based classifier. Their performance for radio detection was also worse compared with that for classifying AGNs.

\section{Conclusions}\label{sec:summary}
Distinguishing AGNs from SFGs for extragalactic radio sources is important to understand the AGN feedback and evolution of galaxies. Existing methods either need costly observations, such as spectroscopy, or cannot satisfy the requirement of purity and completeness simultaneously. In this paper, we provide a supervised AGN classifier based on several large-area surveys, i.e., the LoTSS Deep Field DR1, AllWISE and Gaia DR3. The optical/MIR photometric data from Gaia DR3 and AllWISE are collected to train the model, while the AGN/non-AGN classifications based on SED fitting are treated as the true labels. In order to select the optimal learning algorithms, $F_1$ score and ROC curve are compared among six algorithms. Then the optimal hyperparameters of the best algorithm are determined through the hyperparameter tuning and the stability of model performance. Finally, CatBoost with its default hyperparameters is chosen to build the best classifier.

We also compare model performance for different training features and different datasets. The best performance is achieved when all the optical and MIR magnitude and colors are combined. Although MIR features appear to be more important than optical features, model performance gets worse if only MIR features are considered. Including radio and astrometric features has no improvement on the model performance. The performance with different cross-matched samples confirms that the distributions of train features can also affect the model performance. Uniform model performance can be obtained if the average magnitudes of different training samples are adjusted to be similar.

Our best classifier achieves both high purity and completeness in identifying AGNs, with $precision = 0.974$, $recall = 0.865$ and $F_1 = 0.916$. Compared with AGN classification methods based on color-color diagram and radio/MIR excess, our classifier could improve the completeness through selecting AGN with blue MIR colors and weak radio emission, and improve the purity through excluding non-AGNs with red MIR colors. This classifier is then applied to classify radio sources in the cross-matched sample of LoTSS DR2, AllWISE and Gaia DR3. A sample of 49716 AGNs and 102261 non-AGNs is derived. The reliability of the machine learning based classification is verified by comparing with the spectroscopic classification of SDSS DR17. The purity and completeness for AGN classification can be as high as 94.2 percent and 92.3 percent, respectively. With the training features just from AllWISE and Gaia, our classifier can be easily applied to select and analyze radio AGNs in other radio surveys.


We also build a classifier to identify radio excess sources. The performance is much worse than that for AGN classification, with $precision = 0.714$, $recall = 0.532$ and $F_1 = 0.610$.

\begin{acknowledgements}
We thank the anonymous referee for constructive comments which improve the paper greatly.

This research is funded by National Natural Science Foundation of China (NSFC; grant No. 12003014).

This publication makes use of data products from LOFAR. LOFAR data products were provided by the LOFAR Surveys Key Science project (LSKSP; https://lofar-surveys.org/) and were derived from observations with the International LOFAR Telescope (ILT). LOFAR (van Haarlem et al. 2013) is the Low Frequency Array designed and constructed by ASTRON. It has observing, data processing, and data storage facilities in several countries, which are owned by various parties (each with their own funding sources), and which are collectively operated by the ILT foundation under a joint scientific policy. The efforts of the LSKSP have benefited from funding from the European Research Council, NOVA, NWO, CNRS-INSU, the SURF Cooperative, the UK Science and Technology Funding Council and the Jülich Supercomputing Centre. This publication makes use of data products from the Wide-field Infrared Survey Explorer, which is a joint project of the University of California, Los Angeles, and the Jet Propulsion Laboratory/California Institute of Technology, funded by the National Aeronautics and Space Administration. This work has made use of data from the European Space Agency (ESA) mission
{\it Gaia} (\url{https://www.cosmos.esa.int/gaia}), processed by the {\it Gaia}
Data Processing and Analysis Consortium (DPAC,
\url{https://www.cosmos.esa.int/web/gaia/dpac/consortium}). Funding for the DPAC
has been provided by national institutions, in particular the institutions
participating in the {\it Gaia} Multilateral Agreement. This research has made use of the VizieR catalog access tool, CDS, Strasbourg, France (Ochsenbein 1996). The original description of the VizieR service was published in Ochsenbein et al. (2000).
\end{acknowledgements}

\bibliographystyle{aa} 
\bibliography{bib} 

\end{document}